\documentclass[twocolumn,showpacs,preprintnumbers,prb]{revtex4} 
\usepackage{graphicx}
\usepackage{amsmath}
\usepackage{times}
\usepackage{epsfig}
\usepackage{color}
\usepackage{ulem}   
\begin{document}

\def\salto{\vskip 1cm} \def\lag{\langle} \def\rag{\rangle}
\newcommand{\redit}[1]{\textcolor{red}{#1}}
\newcommand{\blueit}[1]{\textcolor{blue}{#1}}
\newcommand{\magit}[1]{\textcolor{magenta}{#1}}

\title{Many-body calculations of low-energy eigenstates in magnetic
  and periodic systems with self-healing diffusion Monte Carlo: steps
  beyond the fixed phase}

\author{Fernando Agust\'{\i}n Reboredo}     
\affiliation {Materials Science and Technology Division, Oak Ridge
National Laboratory, Oak Ridge, TN 37831, USA}
  
\begin{abstract}
  The self-healing diffusion Monte Carlo algorithm (SHDMC) [Reboredo,
  Hood and Kent, Phys. Rev.  B {\bf 79}, 195117 (2009); Reboredo, {\it
    ibid.} {\bf 80}, 125110 (2009)] is extended to study the ground
  and excited states of magnetic and periodic systems. The method
  converges to exact eigenstates as the statistical data collected
  increases if the wave function is sufficiently flexible. 
  It is shown that the wave functions of complex anti-symmetric
  eigen-states can be written as the product of an anti-symmetric real
  factor and a symmetric phase factor. The dimensionality of the nodal 
  surface is dependent on whether phase is a scalar function or not.
  A recursive
  optimization algorithm is derived from the time evolution of the
  mixed probability density, which is given by an ensemble of
  electronic configurations (walkers) with complex weight.  This
  complex weight allows the amplitude of the fixed-node wave function
  to move away from the trial wave function phase. This novel approach
  is both a generalization of SHDMC and the fixed-phase approximation
  [Ortiz, Ceperley and Martin, Phys Rev. Lett. {\bf 71}, 2777 (1993)].
  When used recursively it simultaneously improves the node and the
  phase.  The algorithm is demonstrated to converge to nearly exact
  solutions of model systems with periodic boundary conditions or
  applied magnetic fields. The computational cost is proportional to
  the number of independent degrees of freedom of the phase. The
  method is applied to obtain low-energy excitations of Hamiltonians
  with magnetic field or periodic boundary conditions.  The method is
  used to optimize wave functions with twisted boundary conditions,
  which are included in a many-body Bloch phase.  The potential
  applications of this new method to study periodic, magnetic, and
  complex Hamiltonians are discussed.
\end{abstract}
\pacs{02.70.Ss,02.70.Tt}
\date{\today}

\maketitle
\section{Introduction} 

Following the basic prescriptions of quantum mechanics, one could
potentially calculate any physical quantity. Finding the solutions of
the Schr\"odinger equation, the wave functions and the associated
energies is all that it is required. However, the computational cost
of obtaining the solutions of many-body problems is well known to
increase exponentially with the number of particles. Minimizing this
exponential cost by either improved algorithms or an insightful
approximation is the central paradigm of condensed matter theory.

Many physical quantities (observables) are only functionals of the
probability density: the square of the modulus of the wave function.
Some other observables, like the ground-state energy, are only 
functionals of the electronic density~\cite{hohenberg}. However, very
important quantities, such as the current density or excitonic
transition matrix elements, depend critically on the wave function.

In confined systems, if the Hamiltonian has time reversal symmetry, a
real-value wave function is well known to exists. However, as soon as
periodic boundary conditions are introduced in the Hamiltonian or a
magnetic field is applied, the amplitude of the wave function is, in
general, complex.  If a wave function has a complex amplitude, both its
modulus $\Phi({\bf R}) $ and its complex phase $e^{{\bf i} \phi({\bf
    R})} $ can depend on the many-body coordinate ${\bf R}=\{{\bf
  r}_1, {\bf r}_2, \cdots , {\bf r}_{N_e} \}$, where ${\bf r}_j$ is
the position of electron $j$ and $N_e$ is the number of electrons.

In addition, for fermions, the many-body wave function must change
sign when the coordinates of any pair ${\bf r}_j$, ${\bf r}_k$ are
interchanged in ${\bf R}$.  In principle, if the wave function is
real-valued, one only needs to determine the exact surface where the wave
function is zero and changes sign (the node) to find the
ground-state energy with diffusion Monte Carlo (DMC) methods.  Any
error in the determination of the node results in an overestimation of
the ground-state energy~\cite{anderson79,reynolds82,HLRbook}.

The standard DMC method~\cite{ceperley80} and improvements~\cite{ortiz93} related to it
require, as an input, a trial wave function $\Psi_T({\bf R}) =
\Phi_T({\bf R}) e^{{\bf i} \phi({\bf R})}$, where both the modulus
$\Phi_T({\bf R})$ and $\phi({\bf R})$ can be chosen to be real.  The
node is the surface $S_T({\bf R})$ in the $3 N_e$ dimensional many-body
space where $\Psi_T({\bf R})= 0$.

The cost of a single DMC step in the standard algorithm is polynomial
in the number of electrons $N_e$, and can be reduced to almost linear
if localized orbitals are used~\cite{williamson,alfe04,reboredo05}. As
a consequence, the existence of an algorithm that finds the required
node with polynomial cost in $N_e$ has been subject of
controversy~\cite{ceperley91,mtroyerprl2005}.  It has been argued that
one of the most important problems in many-body electronic structure
theory is to accurately find representations of the fermion
nodes~\cite{ceperley91,mtroyerprl2005}, which could help in solving the
so-called ``fermion sign problem.''

In general, a guess of the node $S_T({\bf R})$ and the phase
$\phi({\bf R})$ can be obtained from mean field or quantum chemistry
methods [such as density functional theory (DFT), Hartree-Fock,
or configuration interaction (CI)]. This initial trial wave function
is often improved using various
methods~\cite{mfoulkesrmp2001,ortiz95,jones97,guclu05,umrigar07} within a variational
Monte Carlo (VMC) context.  This standard approach depends on the
accidental accuracy of the mean field to find the node or the
possibility to perform accurate CI calculations to pre-select a
multideterminant expansion for $\Psi_T({\bf R})$. In addition, it can
be claimed that a variational optimization of the trial wave function
energy or its energy variance only improves the nodes
indirectly~\cite{luchow07}.

In the last two years, we have developed a method to circumvent the
sign problem for the ground and low-energy eigenstates of confined
systems~\cite{keystone,rockandroll}. This method was recently
validated in real molecular systems~\cite{rollingstones}.  We called
the method self-healing diffusion Monte Carlo (SHDMC) since the nodes
are corrected in a DMC context (as opposed to a VMC optimization) and
the wave function converges to nearly exact~\cite{keystone} or 
state-of-the-art solutions~\cite{rollingstones}, even starting from random.  This
approach is based on the proof~\cite{keystone} that by locally
smoothing the discontinuities in the gradient of the fixed-node ground
state $\Psi_{FN}({\bf R})$ at $S_{T}({\bf R})$, a new trial wave
function can be obtained with improved nodes. This proof enables an
algorithm that systematically moves the nodal surface~\cite{fn:sampling}  $S_{T}({\bf
  R})$ towards that of an eigenstate. The trial
wave function is self-corrected within a recursive DMC approach.  If
the form of trial wave function is sufficiently flexible and given
sufficient statistics, the process leads to an exact eigenstate
many-body wave function~\cite{keystone,rockandroll,rollingstones}.

The success of the fixed-node approximation~\cite{anderson79} used in
the standard DMC algorithm for real wave functions is related to the
quadratic dependence of the error in the fixed-node energy with the
distance between $S_T({\bf R})$ and the exact node~\cite{HLRbook} $S({\bf
  R})$. Because the probability density goes to zero
quadratically at $S({\bf R})$, errors due to small and short 
wave-length departures of $S_T({\bf R})$ from $S({\bf R})$ do not propagate
far into the nodal pocket. Since the DMC energy is dominated by the
average far from the node, DMC tolerates short wave-length departures
of $S_T({\bf R})$ around $S({\bf R})$.

However, if the amplitude of wave function is complex, one must also
determine its phase $\phi({\bf R})$.  The ground-state energy of
complex wave functions can be calculated within the fixed-phase
approximation~\cite{ortiz93} of DMC (FPDMC). But any error in the
phase also results in an overestimation of the ground-state energy
even if the exact nodes are provided~\cite{ortiz93}.  For complex wave
functions, moreover, the error in the phase can be more dramatic than
the nodal error, since the gradient of phase ${\bf \nabla}\phi({\bf
  R})$ is sampled everywhere, and in particular in the regions of
large probability density (see below and Ref. \onlinecite{ortiz93}).

Since (i) periodic or infinite systems are dominant in solid state
physics, (ii) the ability to calculate complex-valued wave functions
(with current) is crucial to understanding transport, (iii) the response
of quantum systems to magnetic fields is key for basic understanding
of correlated phenomena and even applications such quantum
computation, (iv) most physical systems of interest are not confined,
and (v) the error in the phase affects the result more than the
error in the node, solving {\it ``the phase problem''} is,
perhaps, as important as solving the sign problem. 

In this paper a method is derived to simultaneously
obtain not only the node but also the complex amplitude of the trial
wave function for lower energy eigenstates of Hamiltonians with
periodic boundary conditions or under applied magnetic fields.
It is shown that if the phase of the wave function is a scalar function, 
 there is a `{\it special}' gauge transformation 
of the many-body Hamiltonian where the wave functions
is real. These wave functions have nodes that are optimized as in 
original SHDMC method.
 If the phase can only be expressed by multi-valuate function, the nodal surface may have
 a reduced 
dimensionality but there is no constraint to update
the wave-function in SHDMC if the nodes are removed. 

 The
method is applied and validated in a model system studied
previously~\cite{rosetta} where near-analytical solutions can be
obtained.  The scaling of the cost of this new approach is linear in
the number of independent degrees of freedom of the phase. The method
is a generalization of both the ``fixed-phase''
approach~\cite{ortiz93} and the self-healing DMC algorithms developed
to circumvent the sign
problem~\cite{keystone,rockandroll,rollingstones}.  The amplitude of
the wave function is free to adjust to the complex weight of the
walkers in a recursive approach.

A study of
Refs.  \onlinecite{keystone,rockandroll,rollingstones} in reverse chronological order (with increasing detail)  is recommended
before reading this article.  Studying again the seminal fixed-phase
paper by Ortiz, Ceperley and Martin (OCM)~\cite{ortiz93} and the
importance sampling method by Ceperley and Alder~\cite{ceperley80} is
also highly encouraged.

The rest of the paper is organized as follows: In Section
\ref{sc:freephase}, the SHDMC and FPDMC methods are generalized and
blended into a new algorithm that optimizes the complex amplitude (and,
if there is one,
the node) of the trial wave function within a DMC approach. As in the
case of SHDMC, the trial wave function is adjusted recursively within
a generalized DMC approach. In Section
\ref{sc:periodic}, the generalization of SHDMC is
applied to a model Hamiltonian with periodic boundary conditions. The
results are compared with converged CI
results for the same model.  In Section \ref{sc:magnetic}, the Zeeman
splittings of the ground and excited states of a model system are
calculated and compared with converged CI results. Section
\ref{sc:coulomb} describes the results obtained with a realistic
Coulomb interaction. Finally, Section \ref{sc:conclusion} discusses
the advantages, perspectives, and possible applications of these methods
for many-body problems.

\section{A free-amplitude recursive diffusion Monte Carlo method}    
\label{sc:freephase}
This section shows how one can obtain an improved trial wave function
$\Psi_T({\bf R},\tau(\ell +1)) = \langle {\bf R} |\Psi_T^{\ell+1}
\rangle $ by applying a smoothing operator $\hat{D}$ and an evolution
operator $e^{-\tau \hat{\mathcal{H}}_{FN}^{\ell}}$ (during a small
imaginary time $\tau$) to the trial wave function $\Psi_T({\bf R},\tau
\ell) = \langle {\bf R} |\Psi_T^{\ell} \rangle $ provided before. The
limit $\tau^{\prime}=\ell \tau \rightarrow \infty$ is reached
recursively as the iteration index $\ell \rightarrow \infty$.

Following the seminal ideas of OCM~\cite{ortiz93}, $\Psi_T({\bf
  R},\tau^{\prime})$ can be written~\cite{fn:taup} as an explicit
product of a complex phase and an amplitude $\Psi_T({\bf
  R},\tau^{\prime})=\Phi_T({\bf R}) e^{{\bf i} \phi({\bf R})}$.  OCM
chose $\Phi_T({\bf R})$ to be symmetric (bosonic like), real, and
positive, while the phase factor $e^{{\bf i} \phi({\bf R})}$ was antisymmetric
for particle exchanges. 
However, the symmetry of a the phase factor is arbitrary: a symmetric
phase factor can be obtained as
\begin{equation}
\label{eq:phasefactorsymm}
e^{{\bf i} \phi({\bf R})}= \left[ \frac{\Psi_T({\bf R},\tau^{\prime})}{ \Psi_T^*({\bf R},\tau^{\prime} ) } \right]^{1/2}  \; ,
\end{equation}
since both $\Psi_T({\bf R},\tau^{\prime})$ and its complex conjugate change sign for particle exchanges. 
Therefore, any eigenstate can also be  
written as the product of a complex-symmetric phase factor  $e^{{\bf i} \phi({\bf
    R})}$ (like the Jastrow factor) and a real function 
$\Phi_T({\bf R})$ where the symmetry of
$\Phi_T({\bf R})$ depends on whether fermions or bosons are
considered
.  

In this work it is 
proved (see Subsection \ref{ssc:gauge}) that if the phase of a fernionic
eigenstate is a scalar function, then $\Phi_T({\bf R})$ has the same nodal structure
than real functions.
Otherwise $\Phi_T({\bf R})$ might not be zero except for ${\bf r}_i=
{\bf r}_j$.
Thus, the node of the trial wave function $\Psi_T({\bf
  R},\tau^{\prime})$ is given in any case by $\Phi_T({\bf R})$ but the
dimensionality of the nodal surfaces depend on the phase.

The evolution for an additional imaginary time $\tau$ of
$\Psi_T({\bf R},\tau^{\prime})$ is given by
\begin{align}
\Psi_T({\bf R},\tau^{\prime}+\tau) = & \; \label{eq:U}
e^{-\tau \hat{\mathcal{H}}_{FN}^{\ell}} \Psi_T({\bf R},\tau^{\prime}) 
\\ 
= & \; e^{-\tau \hat{\mathcal{H}}_{FN}^{\ell}} \left[  \Phi_T({\bf R})e^{{ \bf i} \phi({\bf R})} \right] \\
= & \; \Phi_T({\bf R},\tau) e^{{ \bf i} \phi({\bf R})}. \label{eq:phi_tau}
\end{align}
Equation (\ref{eq:phi_tau}) includes all the time dependence of the
wave function in $\Phi_T({\bf R},\tau)$, while the phase $\phi({\bf
  R})$ remains fixed~\cite{ortiz93}.

In Eq.~(\ref{eq:U}), $e^{-\tau \hat{\mathcal{H}}_{FN}^{\ell} }$ is the
fixed-node evolution operator, which is a function of the fixed-node
Hamiltonian operator $ \hat{\mathcal{H}}_{FN}^{\ell} $ given by
\begin{equation}
\label{eq:hfn}
\hat{\mathcal{H}}_{FN}^{\ell} = 
   \hat{\mathcal{H}}+ \! \infty \ \lim_{\epsilon \rightarrow 0}
    \theta\left\{\epsilon- d_m[S_T({\bf R'},\ell \tau) - {\bf R}] \right\} \; .
\end{equation}
 The second term on the right-hand side of
Eq.~(\ref{eq:hfn}) adds an infinite potential~\cite{ortiz93} at the
points ${\bf R}$ with minimum distance to any point on the nodal
surface $d_m[S_T({\bf R'},\tau^{\prime})- {\bf R}]$ smaller than
$\epsilon$. The fixed-node Hamiltonian is dependent on $\ell$ since
the nodes $S_T({\bf R'},\tau^\prime)$ change from one iteration to the
next.

In Eq.~(\ref{eq:hfn}), the many-body Hamiltonian $\hat{\mathcal{H}}$ is
given in atomic units by
\begin{equation}
\hat{\mathcal{H}}= \sum_j^{N_e} \frac{
(\nabla_j+{\bf A}_j)^2}{2}+V({\bf R})-E_T 
\end{equation}
where ${\bf A}_j= {\bf A}({\bf r}_j)$ is a vector potential at point
${\bf r}_j$, $V({\bf R})$ includes the electron-electron interaction
and any external potential, 
and $E_T$ is a {\it
  complex}~\cite{fn:gauge} energy reference, adjusted 
to normalize the projected wave function, that cancels out any
phase shift resulting from arbitrary gauge
choices for ${\bf A}_j$ (see remarks below).  

Using Eq~(\ref{eq:U}), one can easily
obtain
\begin{align}
\label{eq:schrodinger} 
\frac{d}{d\tau} \left[ 
 \Phi_T({\bf R},\tau)  
\right]  = & -  e^{-{\bf i} \phi({\bf R})}
 \hat{\mathcal{H}}_{FN}
e^{-\tau \hat{\mathcal{H}}_{FN}}
\Psi_T({\bf R},\tau^{\prime})  \nonumber
\\ 
= & -\left[E_L({\bf R },\tau)-E_T\right] \Phi_T({\bf R},\tau)  
\end{align}
with
\begin{align}
\label{eq:EL}
E_L({\bf R},\tau)  = &   
 -\frac{1}{2}\!\! 
\sum_j^{N_e}  \frac{\nabla_j^2 \Phi_T({\bf R},\tau)}{\Phi_T({\bf R},\tau)}  \\
 & + 
\frac{1}{2} 
 \sum_j^{N_e}
\left|
{\bf A}_j+{\bf \nabla}_j \phi({\bf R})
\right|^2
+V({\bf R}) 
\nonumber \\ 
& - {\bf i} \sum_j^{N_e} \left\{
 \frac{{\bf \nabla}_j  \Phi_T({\bf R},\tau)}{ \Phi_T({\bf R},\tau)}. 
\left[{\bf A}_j \! + \!{\bf \nabla}_j \phi({\bf R})\right]
\right .
  \nonumber
\\ 
& \; \; \; \; \; \; \; \; \; \; \; \; \; + 
\left.  
\frac{{\bf \nabla}_j \cdot \left[ {\bf A}_j 
+ {\bf \nabla_j} \phi({\bf R}) \right]}{2}
\right\}. \nonumber
\end{align}
In Eq.~(\ref{eq:schrodinger}), $ E_L({\bf R },\tau)$ can be a real
constant only if $\Phi_T({\bf R}) e^{{\bf i} \phi({\bf R})}$ is an
eigenstate of $\hat{\mathcal{H}}_{FN}$. In general, for an arbitrary
trial wave function, $ E_L({\bf R },\tau)$ is a complex function of
${\bf R}$. The real part of $ E_L({\bf R },\tau)$ is given by the
first three terms in Eq.~(\ref{eq:EL}), while the imaginary
contribution is given by the last one.  OCM's fixed-phase
approximation results from considering only the real
part~\cite{fn:fixed_phase} of $E_L({\bf R} , \tau )$. With little
effort, one can obtain Eq. (3) of OCM's work assuming $Im[ E_L({\bf R
})]=0$, which leads to a continuity-like equation for fluids.

Note that if $\phi({\bf R})$ is held fixed and $ Im[E_L({\bf R })] \ne 0$ [see
Eqs.~(\ref{eq:U}), (\ref{eq:phi_tau}), (\ref{eq:schrodinger}), and
(\ref{eq:EL})], then $\Phi_T({\bf R},\tau)$ must not only change its
modulus but also must be free to drift away from the real values as
$\tau$ increases.

If at $\tau=0$ an initial distribution of $N_w$ walkers $f({\bf R},0)$
is generated to be equal to $ N_w |\Phi_T({\bf R})|^2$, within a
generalization of the importance sampling algorithm of Ceperley and
Alder~\cite{ceperley80} (see below), $f({\bf R},\tau)$ should evolve
in imaginary time as
\begin{align}
\label{eq:ftau}
f({\bf R},\tau) &= \Phi_T({\bf R}) \Phi_T({\bf R},\tau)  .
\end{align}
Clearly $f({\bf R},\tau)$ can be complex for $\tau > 0$ if
$Im[E_L({\bf R},\tau)]~\ne~0$ [see Eq.~(\ref{eq:schrodinger})].

Replacing Eq.~(\ref{eq:schrodinger}) into Eq.~(\ref{eq:ftau}), and
following a procedure almost identical to the one used in  
Ref.~\onlinecite{ceperley80}, one obtains
\begin{align}
\left. \frac{\partial f({\bf R},\tau)}{\partial \tau}
\right|_{\tau\approx 0}  
 = &
\frac{1}{2}
 \sum_j^{N_e}
\left\{
\nabla_j^2 f({\bf R},\tau) - 
 {\bf \nabla}_j \cdot \left[ f({\bf R},\tau)
{\bf F}_Q^j 
 \right]
\right \}
\nonumber \\
& - \left[ E_L({\bf R})-E_T \right] f({\bf R},\tau) , 
\label{eq:far-dmc}
\end{align}
where
\begin{align}
\label{eq:complexgr}
 {\bf F}_Q^j 
  = {\bf \nabla}_j
ln\left|
\Phi_T({\bf R}) \right|^2 \; ,
\end{align}
and 
$
E_L({\bf R}) = E_L({\bf R},0)
$
is the complex local energy constructed using
Eq.~(\ref{eq:schrodinger}). To obtain Eq.~(\ref{eq:far-dmc}) one
must to assume that 
\begin{equation}
\label{eq:approx}
 \frac{{\bf \nabla}_j  \Phi_T({\bf R},\tau)}{ \Phi_T({\bf R},\tau)} 
\simeq 
\frac{{\bf \nabla}_j  \Phi_T({\bf R})}{ \Phi_T({\bf R})} ,
\end{equation}
which implies that unlike the standard DMC algorithm~\cite{ceperley80},
there is an error in Eq.~(\ref{eq:far-dmc}) when $\tau \rightarrow
\infty$ if $\Psi_T({\bf R},\tau)$ is not an eigenstate. This is only
an apparent limitation since (i) $\tau $ at first can be made as small
as required for Eq.~(\ref{eq:approx}) to be valid, (ii) $\tau $ can be
increased later as the wave function improves and converges to an
eigenstate, (iii) the limit $\tau^{\prime} \rightarrow \infty$ is
reached by applying this free-amplitude method recursively (see
below), and (iv) $\tau$ is already limited to be small in SHDMC with
correlated sampling so that the weights remain close to $1$ (see below).
 
Although Eq.~(\ref{eq:far-dmc}) above for $f({\bf R},\tau)$ is 
identical to Eq. (1) in Ref.~\onlinecite{ceperley80}, it now has a
slightly more {\it complex} interpretation as a stochastic
process. Each member of an ensemble of systems (walker) undergoes (i)
a random diffusion caused by the zero-point motion and (ii) drifting
by the trial quantum force $ ln\left|\Phi_T({\bf R}) \right|^2$
[which depends only on $\Phi_T({\bf R})$ and not on the
phase], but in variance with Ref. \onlinecite{ceperley80}, (iii)
each walker carries a complex phase. In a nonbranching algorithm, the
complex weight of the walkers is multiplied by $\exp\{-\left[ E_L({\bf
    R})-E_T \right]\delta\tau\}$ at every time step.

Similar to the case of the ``simple'' SHDMC algorithm (see
Refs.~\onlinecite{keystone,rockandroll,rollingstones} for details),
the weighted distribution of the walkers can be written as
\begin{align}\label{eq:evoltau}
  f({\bf R},\tau) 
                & =  \lim_{N_c \rightarrow \infty} \frac{1}{N_c} \sum_{i=1}^{N_c}
   W_i^j(k) \delta \left({\bf R-R}_i^j \right). 
\end{align}
In Eq.~(\ref{eq:evoltau}), ${\bf R}_i^j$ corresponds to the position
of the walker $i$ at step $j$ of $N_c$ equilibrated
configurations. The {\it complex} weights $W_i^j(k)$ are given by
\begin{align}
\label{eq:weights}
W_i^j(k) = e^{-\left[
E_i^j(k)-E_{T}
\right] \tau}
\end{align}
with 
\begin{align}
  E_i^j(k) = \frac{1}{k}\!\sum_{\ell=0}^{k-1} E_L({\bf R}_i^{j-\ell}),
\end{align} 
where $E_T$ in Eq.~(\ref{eq:weights}) is now a {\it complex} energy reference 
periodically adjusted so that
$\sum_i W_i^j(k)\approx N_c$ and $\tau$ is $k\delta\tau$ ($k$
is a small number of steps and $\delta\tau$ is a standard DMC time step).

The trial wave function $\Psi_T({\bf R},\tau^{\prime}+\tau)$ for the
next iteration can be obtained as follows:  All wave functions
can be expanded in a basis as
\begin{align}
\label{eq:trialwf}
\Psi_T({\bf R},\tau^{\prime}) & = 
 e^{J({\bf R})} 
\sum_n^{\sim} \lambda_n(\tau^\prime)  \Phi_n({\bf R}) 
.
\end{align}
In Eq. (\ref{eq:trialwf}), $\sum_n^{\sim}$ represents a truncated sum,
$\{\Phi_n({\bf R})\}$ forms a complete orthonormal basis of the
antisymmetric Hilbert space~\cite{fn:basis}, and $e^{J({\bf R})}$ is a
symmetric Jastrow factor. The $\lambda_n(\tau^\prime)$ are complex
coefficients to be defined [see Eq. (\ref{eq:lambda})]. Note that 
the expressions~\cite{fn:taup}
\begin{align}
\label{eq:defPhi}
\Phi_T({\bf R}) & = \pm \sqrt{\Psi_T({\bf R},\tau^{\prime})\Psi_T^*({\bf R},\tau^{\prime})   }, \text{    and } \\
\label{eq:phaselog}
\phi({\bf R}) &= \ln[\Psi_T({\bf R},\tau^{\prime})/\Psi_T^*({\bf R},\tau^{\prime}) ]/ (2 i) +  \pi n 
\end{align}
allow the computation of all the quantities involved in $E_L({\bf R})$
in terms of gradients and Laplacians of $\Phi_n({\bf R})$ and $J({\bf
  R} )$.  In Eq. (\ref{eq:phaselog}) $n$ is an arbitrary integer that
changes the Riemann branch of the natural logarithm $\ln$, but does not contribute
to the gradient within a branch. The local energy is thus independent on the choice of $n$
but at the Riemann cuts where, sometimes, $n$ has to change to make the
phase continuous. However, the probability of a walker to touch the Riemann 
cut is, in practice, zero. 

From Eqs.~(\ref{eq:U}),  (\ref{eq:ftau}) and  (\ref{eq:evoltau}), one can
formally obtain
\begin{align}
\label{eq:nextevol}
 \tilde \Psi_T({\bf R},\tau^{\prime}+\tau)  = & \;  e^{{\bf i}\phi({\bf R})} f({\bf R},\tau^{\prime}+\tau)/ 
 \Phi_T({\bf R}) \\
  =& \; \langle {\bf R} | e^{-\tau \hat{\mathcal{H}}_{FN} }|\Psi_T(\tau^{\prime}) \rangle \; . 
\end{align}
The local smoothing operator is defined as 
\begin{align}
\label{eq:deltaexp}
\langle {\bf R}^{\prime}| \hat{D} |  {\bf R} \rangle & = \tilde \delta \left( {\bf R^{\prime},R} \right)  \\
 & =
\sum_n^{\sim}
e^{J({\bf R^{\prime}})} \Phi_n({\bf R^{\prime}}) \Phi_n^*({\bf R}) e^{-J({\bf R})} 
\nonumber
.
\end{align}
Applying Eq.~(\ref{eq:deltaexp}) to both sides of Eq.~(\ref{eq:nextevol}), 
using Eq.~(\ref{eq:evoltau}),
and integrating over ${\bf R}$, 
one can easily obtain
\begin{align}
\label{eq:sequence}
\Psi_T({\bf R},\tau^{\prime}+\tau) 
= & \;\langle {\bf R} | \hat{D} e^{-\tau \hat{\mathcal{H}}_{FN} } | \Psi_T(\tau^{\prime}) \rangle \\
= & \; e^{J({\bf R})} 
\sum_n^\sim \langle \lambda_n(\tau^\prime+\tau) \rangle \Phi_n({\bf R}) ,
\end{align}
with 
\begin{align}
\label{eq:lambda}
\langle \lambda_n(\tau^\prime\!+\!\tau) \rangle=
 \frac{1}{\mathcal{N}} \sum_i^{N_c} 
 W_i^j(k)  e^{-J({\bf R}_i^j)} \frac {\Phi^*_n ({\bf R}_i^j)}
 { \Psi^*_T ({\bf R}_i^j,\tau^\prime)}\gamma({\bf R}_i^j)
\end{align}
where $ \mathcal{N}= \sum_{i=1}^{N_c} e^{-2J({\bf R}_i^j)} $
normalizes the Jastrow factor. 
$\gamma({\bf R}) $ is the standard time-step correction [Eq.~(33) in
Ref.~\onlinecite{umrigar93}]: 
\begin{align}
\label{eq:gamma}
 \gamma ({\bf R})= \frac{-1 + \sqrt{1 + 2 |{\bf v}|^2 \tau}}
{|{\bf v} |^2  \tau} 
\text{ with }
{\bf v} = \frac{\nabla \Phi_T({\bf R})} {\Phi_T({\bf R})}.
\end{align}

Note that $ \Phi_T ({\bf R})$ includes a Jastrow factor, thus
Eq.~(\ref{eq:deltalambda}) reduces to the one used in the original 
``simple'' SHDMC
algorithm~\cite{keystone,rockandroll} for $\phi({\bf R})= 0$.

In addition, as suggested by
Umrigar~\cite{umrigar_private} for the ground-state SHDMC
algorithm~\cite{keystone}, correlated sampling can be used also for
walkers with complex weight. 
One can sample $\delta\lambda_n =
\lambda_n(\tau^\prime+\tau)-\lambda_n(\tau^\prime)$, which
results in
\begin{align} 
\label{eq:deltalambda}
& \langle \lambda_n(\tau^\prime+\tau) \rangle = \lambda_n(\tau)+
\langle \delta \lambda_n \rangle 
\\ \nonumber & \langle \delta
\lambda_n \rangle = \frac{1}{\mathcal{N}} \sum_{i=1}^{N_c} e^{-J({\bf
    R}_i^j)} \frac {\Phi_n^* ({\bf R}_i^j)} 
{ \Psi^*_T ({\bf    R}_i^j,\tau^\prime)} [W_i^j(k)-1] \gamma({\bf R}_i^j).
\end{align}
  These new $\lambda_n(\tau^\prime+\tau)$
[Eq.~(\ref{eq:deltalambda})] are used to construct a new trial wave
function [Eq.~(\ref{eq:trialwf})] recursively within DMC. Equation
(\ref{eq:lambda}) can be related to the maximum-overlap method used
for bosonic wave functions~\cite{reatto82}.

The error of $\langle \delta \lambda_n \rangle $ is obtained by sampling
\begin{align} 
\label{eq:deltalambda2}
& 
\langle \delta \lambda_n^2 \rangle  =  
\frac{1}{\mathcal{N}} \sum_{i=1}^{N_c} \left| 
e^{-J({\bf R}_i^j)} 
\frac {\Phi_n^* ({\bf R}_i^j)} 
{ \Psi^*_T ({\bf R}_i^j)} 
[W_i^j(k)-1] \gamma({\bf R}_i^j)
\right|^2 .
\end{align}
The truncation of the expansion of the delta function
[Eq.~(\ref{eq:deltaexp})] is a key ingredient in SHDMC since it
decides how local is the smoothing operator $\hat{D}$ and prevents
noise to ruin the quality of the trial wave function.  If the absolute
value of the error of $\langle \delta \lambda_n \rangle $ is larger
than $|\lambda_n(\tau^\prime+\tau)|/4$, the algorithm sets
$\lambda_n(\tau^\prime+\tau)$ equal to zero [which defines the
truncation criterion used in the sums ($\sum_n^{\sim}$) involved in
Eqs.~(\ref{eq:trialwf}), (\ref{eq:deltaexp}) and (\ref{eq:sequence})].
See Ref. \onlinecite{keystone} for a detailed theoretical
justification of the truncation procedure and the algorithm used.
Briefly here, the coefficients $\lambda_n$ are sampled at the end of
each sub-block of $k$ DMC steps. Statistical data is collected for
number of sub-blocks $M$ before a wave function update. At first, $M$
is set to a small number and increased according to the recipe given
in Ref. \onlinecite{rockandroll}.  In short, the algorithm detects
automatically the dominance of noise when the projection of two
successive sets of $\langle \delta\lambda_n \rangle$ becomes small and
multiplies by a factor larger than 1 the number of sub-blocks $M$ (see
Ref. \onlinecite{rockandroll} for more details).  As a result, the
total number of configurations $N_c$ sampled increases as the
algorithm progresses. Therefore the statistical error is reduced, and
the number of basis functions retained in the expansion increases over
time. Thus, the smoothing operator $\hat{D}$ tends to the delta
function as $M$ increases, which allows the SHDMC method to sample the
wave function with increasing detail.  The first quarter of the data
in each block, following a wave function update, is discarded.

In practice, the only difference between this new approach and the
original SHDMC method is the complex weight and the limitation for
propagation to small $\tau$. Therefore, a nonbranching algorithm for
small $\tau$ has been used (see Ref. \onlinecite{rockandroll} for
details). However, there are some formal differences on the
justification of the convergence of the SHDMC method that are
discussed in the next subsections.

\subsection{Gauge transformations and nodal structure of complex eigenstates} 
\label{ssc:gauge}

The phase $\phi({\bf R})$ must be continuous at any point of ${\bf R}$
where $\Psi_T({\bf R}) \ne 0$.  Otherwise, if $\phi({\bf R})$ is not
continuous, its gradient in the local energy will introduce an
effective infinite potential at the discontinuity that will force
$\Psi_T({\bf R}) = 0$.  In some cases, however, these discontinuities in
the phase are not physical\cite{fn:discont} and they
can be removed by changing the Riemann sheet index $n$ in
Eq. (\ref{eq:phaselog}).  As a consequence, the wave functions
can be split into two classes.  In the first class different
Riemann sheets of $\phi({\bf R})$ are {\it not} connected. In that
case, one can choose as phase a single sheet of the Riemann surface 
The phase in this class can be a continuous scalar
function of ${\bf R}$ for every ${\bf R}$. The real wave functions,
with constant phase, are special case of this class.  In the second
class, the Riemann surfaces of $\phi({\bf R})$ for different $n$ are connected
at the Riemann cuts. Thus a continuous $\phi({\bf R})$ can only be described by
 a multi-valuate function of ${\bf R}$.

{\it Eigenstates with a scalar phase:} 
Since ${\bf \nabla}_j \times {\bf \nabla}_j\cdot \Lambda({\bf R}) = 0$
for any scalar function $\Lambda({\bf R})$, the magnetic field ${\bf
  B} = {\bf \nabla}_j \times {\bf A }_j $ is invariant for the gauge
transformations~\cite{jackson} ${\bf A }^{\prime}_j= {\bf A }_j+ {\bf
  \nabla}_j [\Lambda({\bf R})+c(\tau)]$, where $\Lambda({\bf R})$ is
an arbitrary symmetric scalar function of ${\bf R}$ and $c(\tau)$ is
an arbitrary function of $\tau$ (independent of every ${\bf r}_j$ in
${\bf R}$).  If $\Psi_T({\bf R} )$ is selected to be an eigenstate of
$\hat{\mathcal H}$ for a given gauge and if $\phi({\bf R})$ is a
scalar function , then a change in gauge ${\bf A }_j+ {\bf \nabla}_j
\delta\Lambda({\bf R}) $ could be readily compensated in the phase by
\begin{equation}
\label{eq:gauge}
\tilde\phi(\bf R) = \phi({\bf R})-\delta\Lambda({\bf R})+ c(\tau),
\end{equation}
without affecting $\Phi_T({\bf R})$ [since ${\bf A }_j $ and $ {\bf
  \nabla}_j \phi({\bf R}) $ always appear added in Eq.
({\ref{eq:EL}})].  This property is particularly important, since implies that 
for this class of eigenstates of $\hat{\mathcal H}$ there is a `{\it special}' gauge 
where the wave function is real.

Note that if one sets $\delta\Lambda({\bf R})=\phi({\bf R})$ in 
Eq. (\ref{eq:gauge}) then
$\tilde\phi(\bf R)~=~0$. Therefore, the new phase is a constant
that can be chosen to be zero. 
The vector potential in this special gauge is a many-body object which includes the gradient of
the many-body phase of the wave function in a single particle gauge.

The norm $\Phi({\bf R})$ is invariant since the effective potential
in Eq. (\ref{eq:EL}) is invariant using Eq. (\ref{eq:gauge}).  This is
expected since the expectation value of an arbitrary operator
$\hat{\mathcal{O}}({\bf R})$ must be independent of the gauge choice
for non-degenerate eigenstates.
In particular, the nodes, which are given by $\Phi({\bf R})$, are also
invariant to gauge transformations.  Since the new phase is a constant, it
can be easily shown that in the special gauge, the amplitude of those
eigenstates has the same structure as the trial wave function used in
the fixed-node approximation for real wave functions.
 
SHDMC self adjusts to an arbitrary gauge change $\delta\Lambda({\bf R})$ because
$\phi({\bf R})$ is modified recursively by a change in the local energy in Eq.
(\ref{eq:EL}) of the form
\begin{align}
\label{eq:deltaEL}
&\delta E_L({\bf R},\tau)  =  \\
& 
 \sum_j^{N_e} \left\{
Re\left[\left(
{\bf A}_j+{\bf \nabla}_j \phi({\bf R})
\right) \cdot{\bf \nabla_j} \delta\Lambda({\bf R}) \right]  + \frac{1}{2} |{\bf \nabla}_j \delta\Lambda({\bf R})|^2 \right\}
\nonumber \\ 
&  - \;  {\bf i} \sum_j^{N_e} \left\{
 \frac{{\bf \nabla}_j  \Phi_T({\bf R},\tau)}{ \Phi_T({\bf R},\tau)}\cdot {\bf \nabla_j} \delta\Lambda({\bf R}) 
 +  
\frac{1}{2}\nabla^2_j \delta\Lambda({\bf R} )
\right\}. \nonumber
\end{align}

{\it Eigenstates with a multi-valuate phase:}
On the other hand if $\phi({\bf R})$ is not a scalar function then,
${\bf \nabla}_j \times {\bf \nabla}_j\cdot \phi({\bf R}) \ne 0$ and,
therefore, $ {\bf \nabla}_j\cdot \phi({\bf R})$ cannot be included in
the vector potential without introducing an artificial many-body
magnetic field. In that case, as pointed out timely by an
anonymous referee, there might
be nodes only where ${\bf r}_j = {\bf r}_k$. Eigenstates with this
type of nodes, with reduced dimensionality, can be found in states
with current, degeneracy or magnetic fields.

Sumarizing, the norm of the complex wave-functions of the eigenstates
that have a scalar phase in ${\bf R}$ has the same structure as the
real wave function used in the fixed-node approach because there is a
special gauge transformation where the wave function is real valued.
This property has a formal importance since it allows extending
theorems developed in the context of the fixed-node approximation.  Instead, 
if different Riemann sheets of $\phi({\bf R})$ are continuously connected, a continuous
phase cannot be described by a single scalar function. Then
dimensionality of the nodal surface might be smaller and limited to
the cases in ${\bf R}$ where ${\bf r}_j ={\bf r}_k$.  The nodes are an
obstacle for DMC; SHDMC, however, converges to eigenstates regardless
of the dimensionality of the nodes (see below).

\subsection{Remarks on the free-amplitude SHDMC method}

{\it Convergence of SHDMC to eigenstates:} In \ref{ssc:gauge} it is
shown that the wave-function of any fermionic eigenstate can be
factorized into an anti-symmetric real function $\Phi_T({\bf R})$,
with nodes, and a symmetric phase factor (See Eq.
(\ref{eq:phasefactorsymm})). The dimensionality of the nodal surface
depends on the phase properties.  Since the amplitude can not change at
the node in DMC, nodes are the obstacle to overcame by SHDMC.
Convergence is not affected if the initial trial wave function has no
nodes because, in this case, the amplitude and the phase can evolve
with $\tau$ everywhere.  Indeed, SHDMC can be started from a linear
combination of real and imaginary parts with different nodes. As noted
by a referee, in this class of functions two particles can exchange
without crossing a node. However, the convergence of SHDMC is not
affected in theory and it is not affected in practice. 
But, if the phase is a scalar function, 
nodes will develop as the trial wave function
converges to an eigenstate. In that
limit, a kink at the node should appear\cite{keystone} until 
the exact node is found.

The convergence of the SHDMC approach when the trial wave function
approaches an eigenstate and shows nodes, is based on the
proof~\cite{keystone} that locally smoothing the kinks of the
fixed-node wave function improves the nodes.  This proof can be
trivially extended to a complex wave function (when $Im[ E_L({\bf R
})]\ne 0$~), breaking the time evolution into a sequence of pure real
evolution followed by an imaginary evolution. If one assumes $Im[
E_L({\bf R })]=0$, the present approach reduces to SHDMC method, but
using the effective potential of the fixed-phase
Hamiltonian~\cite{ortiz93}.  Thus the best nodes in $\Phi_T({\bf
  R},\tau)$ for a given phase $\phi({\bf R})$ can be obtained by
running SHDMC in a fixed-phase stage. It is trivial to show that the
phase, in turn, improves if the imaginary contribution is allowed to
evolve a short time from a trial wave function with optimal nodes. In
principle, a SHDMC fixed-phase stage can be propagated to infinite
imaginary time $\tau$ within a branching algorithm. The evolution of
the phase, instead, is limited to short times [for
Eq.~(\ref{eq:approx}) to be valid].  In this work, however, the real
and the imaginary parts of the wave function are allowed to evolve
simultaneously during a short time without any observed adverse effect
on the accuracy.

{\it Phase and nodal errors:} Note in Eq.~(\ref{eq:EL}) that as in the
fixed-phase approach~\cite{ortiz93}, an effective potential $\left|
  {\bf \nabla}_j \phi({\bf R})+ {\bf A}_j \right|^2$ is added to
$V({\bf R})$, which depends in turn on ${\bf \nabla} \phi({\bf R})$
and ${\bf A}_j$. Thus small errors in the phase $ \phi({\bf R})$ have
a global impact [in particular far from the node, where $|\Phi_T({\bf
  R})|^2 >> 0$]. In contrast, small errors in $\Phi_T({\bf R})$ only
slightly displace the node and have smaller impact on the
energy~\cite{HLRbook} [since the local energy is seldom sampled
because $\lim \Phi_T({\bf R})\Phi_T({\bf R},\tau) \rightarrow 0$ at
$S_T({\bf R})$]. For eigenstates without nodal surfaces the phase is
the sole source of error.  SHDMC provides a method to correct both the
phase and the nodal error.

{\it Complex $E_T$:} 
Note that when $c(\tau)$ is changed in Eq. (\ref{eq:gauge}), it just
changes the normalization of $\Psi_T({\bf R})$ and, if complex, introduces a
global phase shift; however, $c(\tau)$ does not affect the local
energy or other observables.

For a random trial wave function $\Psi_T({\bf
  R})$ and an arbitrary choice of gauge for ${\bf A}_j$, $E_L({\bf
  R},\tau)$ will have both real and imaginary components. The average
over the walkers` positions will have a real contribution, which affects
the norm, and a complex contribution, which introduces a global phase
shift. The average of $E_L({\bf  R},\tau)$ only contributes to $c(\tau)$.  The correlated
sampling approach is obviously more efficient when the
distribution of complex weights is centered around 1 since the error in the coefficients
is minimized [See Eq. \ref{eq:deltalambda2} and use 
the standard expression for the variance].  $E_T$ in
Eq.~(\ref{eq:weights}) is thus complex.  The real part of $E_T$
renormalizes the wave function (that is, keeps the population of
walkers constant). The imaginary part of $E_T$ removes the global
phase shift [the average complex contribution of $E_L({\bf R},\tau) $
that only contributes to $c(\tau)$]. For a converged trial
wave-function, $Im(E_T) \simeq 0$.

{\it Upper bound properties:} The present approach should not be
considered as a method to estimate the energy of the trial wave
function but instead as a method to optimize the trial wave function
before a final FPDMC calculation.  However, since the real part of
$E_L({\bf R})$ corresponds to the fixed-phase approximation, the real
part of $E_T$ also converges to an upper bound of the ground-state
energy. This upper bound can be higher than the fixed-phase
approximation (if the limit of $\tau^{\prime} \rightarrow \infty$ is
not reached or the basis $\{\Phi_n({\bf R})\}$ is too small).
Therefore, a standard FPDMC calculation~\cite{ortiz93} (with
branching) must be performed to obtain final values for the
ground-state energy.
 
{\it Known limitations and solutions:} The present approach is inefficient
when the energy of the first excited state $E^{FP}_1$ of the
fixed-phase Hamiltonian is too close to the ground-state energy
$E^{FP}_{0}$, since the coefficient of the excited state component of
the trial wave function decays as $exp[-\tau(E^{FP}_1-E^{FP}_{0})]$.
In that regime, satisfactory results can be obtained as follows: begining with a small
value for $M$, keep $M$ fixed for a number of iterations until the
lower energy excitations decay and, then release $M$, allowing the
high energy components of the wave function to converge.

Moreover, the approximate excited-state wave
functions can be calculated (see Ref. \onlinecite{rockandroll}) and the lowest
energy linear combination can be determined with correlated function Monte Carlo
\cite{jones97}, VMC~\cite{umrigar07} or directly in FPDMC using a
restricted basis of low energy states. A final alternative is to run
this free-amplitude SHDMC method with larger $\tau=k \delta\tau$ but
using a smaller basis given by a few approximated excited states. The
excited states can be found as described in the next section.

\subsection{Generalization  to excited states}
\label{ssc:excited}
Earlier estimates of excited state energies in the presence of
magnetic fields have been made by diagonalizing a matrix of correlation
functions in imaginary time~\cite{jones97,ceperley88}. In addition, 
calculations of excited states have been reported with the 
auxiliary field approach~\cite{purwanto09}. The present
algorithm, in contrast, is almost identical to the SHDMC
method~\cite{keystone,rockandroll} developed for the ground state and
lower excitations of real wave functions. The only relevant difference
with Ref. \onlinecite{rockandroll}
is the complex weight of the walkers. Thus the free-amplitude SHDMC
method described above for the ground state can be generalized in a
straightforward way to study excited states as in Ref.
\onlinecite{rockandroll}.

Readers are encouraged to follow a detailed theoretical
justification of the excited state algorithm in
Ref.~\onlinecite{rockandroll}. Here only some key steps are described
[in particular, note Eqs.~(\ref{eq:bra}) and (\ref{eq:xi}) that were omitted
in Ref.~\onlinecite{rockandroll} and are relevant for a non-unitary
Jastrow factor].

As in the importance sampling algorithm~\cite{ceperley80}, the
generalization given by Eq.~(\ref{eq:far-dmc}) requires that
$\Psi_T({\bf R},\tau^\prime\; +\;\tau)$ be zero only at the nodes
$S_{T}({\bf R},\tau^\prime)$ of $\Psi_T({\bf R},\tau^\prime)$, being
free to change both its modulus and phase elsewhere. Therefore,
$\Psi_T({\bf R},\tau^\prime +\tau)$ can develop a projection into any
lower energy state consistent with $S_{T}({\bf R},\tau^{\prime})$.  To
obtain an excited state, the wave function $\Psi_T({\bf R},\tau^\prime
+\tau)$ must be projected in the subspace orthogonal to the ground
state and any other excited state calculated before. In alternative
approaches such as correlation function Monte Carlo~\cite{ceperley88},
the orthogonality of excited states is achieved by diagonalizing a
generalized eigenvalue problem.  One could argue that the excited
states obtained with that approach share nodal error of the ground
state.  One of the advantages of SHDMC is that the diagonalization of
a large matrix of excitations is avoided, which makes possible the
consideration of a larger number of degrees of freedom. In addition, 
the nodes of each excitations are found independently.  But in SHDMC,
unless special conditions are satisfied~\cite{rockandroll}, one must
calculate lowerlying energy states before attempting the calculation
of higher excited states.

A projector is constructed with approximated expressions of the $\nu$
eigenstates $\Psi_{\mu}({\bf R})= \langle {\bf R}| e^{\hat{J}} \breve
\Phi_{\mu} \rangle =e^{J({\bf R})}\breve \Phi_{\mu}({\bf R})$
calculated earlier as
\begin{equation}
\label{eq:orthogonality}
\hat{P}_\nu = e^{\hat{J}}\left[ 1 - \sum_{\mu}^{\nu} 
|\breve \Phi_{\mu}\rangle \langle\breve \Phi_{\mu}^\dagger|\right]   e^{-\hat{J}} \; \;.
\end{equation}
The operator $e^{\hat {J}}$ in Eq.~(\ref{eq:orthogonality}) is 
the multiplication by a Jastrow. For a non-unitary  $e^{\hat {J}}$ the 
set $\{ | \breve \Phi_{\mu}\rangle  \}$ is nonorthogonal. However, the
conjugate (dual) basis~\cite{hoffman,prugovecki} that satisfies $\langle\breve \Phi_{\mu}|\breve \Phi_m\rangle =
\delta_{\mu,m}$ can be obtained statistically as 
\begin{equation}
\label{eq:bra}
\langle \breve \Phi_\mu |{\bf R} \rangle  = \sum_n^\nu \xi_n^\mu  \Phi^*_n ({\bf R})
\end{equation}
with
\begin{align}
\label{eq:xi}
\xi_n^\mu =
\lim_{
N_c \rightarrow \infty}
\frac{1}{\mathcal{N}} \sum_i^{N_c} 
\frac{W_i^j(k)}{e^{-J({\bf R}_i^j)}}  \frac {\Phi_n ({\bf R}_i^j)} 
 { \Psi_T^{\mu } ({\bf R}_i^j,\tau^\prime)} \gamma ({\bf R}_i^j) \;,
\end{align}
where $\Psi_T^{\mu} ({\bf R},\tau^{\prime})$ is the trial
wave function used to evaluate earlier the state $\mu$ for
$\tau^{\prime} \rightarrow \infty $.  Note that the exponential involving
$-J({\bf R}_i^j)$
moves to the denominator in Eq.~(\ref{eq:xi}) as compared with
Eq.~(\ref{eq:lambda}). Since $J({\bf R})$ is real, the phase of
$\langle \breve \Phi_\mu |{\bf R} \rangle $ must be conjugated to the
phase of $\langle {\bf R} | \breve \Phi_\mu \rangle$. The
coefficients $\xi_n^\mu$ should be sampled during the final FPDMC step
(i.e., when the final excited energy is sampled).

The projection of the conjugate function $\langle
\breve \Phi_\mu |$ onto earlier conjugate states should also be
removed to obtain $\langle \breve \Phi_\mu |= \langle \breve \Phi_\mu
| \hat{P}^T_{\mu-1}|$ where $\hat{P}^T_{\nu}$ is the transpose of
$\hat{P}_{\nu}$.  Furthermore, statistical errors in 
$\langle\breve \Phi_{\mu}|$ can be partially filtered by inverting the
overlap matrix $S_{\mu,m}= \langle\breve \Phi_{\mu}|\breve
\Phi_m\rangle$ as
\begin{equation}
  \langle\breve \Phi_{\mu}^\dagger|= \sum_m S^{-1}_{\mu,m} \langle\breve \Phi_m| \; .
\end{equation}

The scalar products resulting from applying
 $\hat{P}_\mu$ in Eq.~(\ref{eq:orthogonality}) are given by
\begin{equation}
\label{eq:scalar}
\langle\breve \Phi_{\mu}^\dagger|\breve \Phi_m\rangle = \sum_n \bar{\xi}_n^\mu \lambda_n^m ,
\end{equation}
with $\bar{\xi}_n^\mu = \sum_\nu S^{-1}_{\mu,\nu} \xi_n^\mu  $
since $\int \Phi_m^* ({\bf R}) \Phi_n ({\bf R}) {\bf dR}=
\delta_{n,m}.$

The extension of SHDMC to the next excited $|\Psi_{\nu+1}\rangle$
can be thought of as the recursive application of the evolution operator
$e^{-k \delta\tau \hat{\mathcal{H}}^{(\ell-1)}_{FN}}$, the projector
$\hat{P}$ [Eq~(\ref{eq:orthogonality})], and a smoothing operation
$\hat{D}$ [see Eq.~(\ref{eq:deltaexp})] to a trial wave function 
$|\Psi_{T,\nu+1}^{\ell-1}\rangle$
[see Eq.~(\ref{eq:SHDMCexcited})].
 This
procedure can be derived analytically~\cite{rockandroll} as follows:
\begin{eqnarray}
\label{eq:SHDMCexcited}
\nonumber
|\Psi_{\nu+1}\rangle & = & 
\lim_{\tau \rightarrow \infty} \hat{P} \; e^{-\tau \hat{\mathcal{H}} }\hat{P}  
|\Psi_{T,\nu+1}^{\ell=0}\rangle \\
           \nonumber
         & = &  
\lim_{\ell \rightarrow \infty} 
\hat{P} \;
\prod_{\ell} 
\left( 
 e^{-(\delta \tau^{\prime}+k \delta\tau ) \hat{ \mathcal{H}}} \hat{P}
\right) 
| \Psi_{T,\nu+1}^{\ell=0}\rangle
\\
           \nonumber
         & = &  
\lim_{\ell \rightarrow \infty} 
\hat{P} \;
\prod_{\ell} 
\left( 
 e^{-\delta\tau^{\prime}  \hat{\mathcal{H}}}  
 e^{-k \delta\tau  \hat{\mathcal{H}}_{FN}^{(\ell-1)}} 
\hat{P}
\right) 
| \Psi_{T,\nu+1}^{\ell=0}\rangle
 \\ 
         & \simeq &  
\lim_{\ell \rightarrow \infty} 
\hat{P}
\prod_{\ell} 
\left( 
 \tilde D e^{-k \delta\tau  \hat{\mathcal{H}}^{(\ell-1)}_{FN}} \hat{P}
\right) 
| \Psi_{T,\nu+1}^{\ell=0}\rangle \\ \nonumber
         & = & | \Psi_{T,\nu+1}^{\ell \rightarrow \infty }\rangle.
\end{eqnarray}
Replacing $ e^{k \delta\tau  \hat{ \mathcal{H}}} $ in the infinite 
product in the second line 
of Eqs. (\ref{eq:SHDMCexcited}) 
with
$e^{-k \delta\tau \hat{\mathcal{H}}^{(\ell-1)}_{FN}}$ 
in the third line generates the same projector~\cite{keystone,rockandroll}.
 In turn, we proved~\cite{keystone} that replacing
 $e^{-\delta\tau^{\prime} \hat{\mathcal{H}}}$ with a large class of
 local smoothing operators $D$ has the same effect on the nodes.
The fixed-node Hamiltonian depends on the iteration index $\ell$ because the 
 trial wave function, the node, $E_T$, and the phase are different at every iteration.
 Finally, the norm of the projected function can be fixed by adjusting
 $E_T$ in every iteration $\ell$.
 
 For states with inequivalent nodal pockets, special care must be
 taken within the algorithm to avoid systematic errors (see
 Ref.~\cite{rockandroll} for additional details about the algorithm).
\section{Calculations for Hamiltonians with  periodic boundary conditions}
\label{sc:periodic}

Usually periodic boundary conditions in a supercell with dimensions
$a_x$, $a_y$ and $a_z$ are set when studying crystalline systems that simulate
an infinite solid.  By using the Bloch Theorem~\cite{ashcroft}, the trial
wave function at $\tau^\prime=\ell \tau$ can be written as the product
of a many-body phase\cite{rajagopal95} times a periodic part~\cite{fn:taup}
$\mathcal{U}({\bf R})$ as
\begin{equation}
\label{eq:periodic}
\Psi_T({\bf R},\tau^\prime) = 
e^{ {\bf i} \left(\sum_j^{N_e} {\bf k\cdot r}_j\right) }
\mathcal{U}({\bf R})
\end{equation}
with 
\begin{align}
\mathcal{U}(\{{\bf r_1},\cdots,{\bf r}_j,\cdots,{\bf r_{N_e}} \})=
\mathcal{U}(\{{\bf r_1},\cdots,{\bf r}_j+{\bf a},\cdots,{\bf r_{N_e}} \})
\nonumber
\end{align}
for any $j$, where ${\bf a}= a_x n_x {\bf \hat{\i}}+ a_y n_y {\bf
  \hat{\j} } + a_z n_z {\bf \hat{k} } $, with $n_{\mu}$ being arbitrary
integers. $\mathcal{U}({\bf R})$, in turn,~\cite{fn:taup} can be
written as a product of a multi-determinant expansion times a Jastrow
factor. Each orbital entering each determinant in $\mathcal{U}({\bf
  R})$ can be expanded in plane waves that satisfy periodic boundary
conditions.

The theory developed in Section \ref{sc:freephase} can then be applied
to periodic systems by setting ${\bf A}_j=0$.  Note, however, that since
$\mathcal{U}({\bf R})$ is in general a complex function, the phase
entering in $E_L({\bf R})$ [see Eq.~\ref{eq:EL}] must include both the
phase of $\mathcal{U}({\bf R})$ and the many-body Bloch phase. The
resulting wave function $\Psi_T({\bf R},\tau^\prime \rightarrow
\infty)$ corresponds, in general, to a state with current, and its
solution can facilitate the calculation of transport problems in a
many-body context~\cite{krcmar08}. The many-body Bloch phase component on
the trial wave function is often referred to in the literature as twisted
boundary conditions~\cite{tbc,fn:tbc}.

\subsection{The model periodic system}

Until this new development, the author had promised himself to halt
calculations in small model
systems~\cite{rosetta,keystone,rockandroll}. Those small model
systems, however, while not very realistic, allow 
comparisons to be performed with fully converged CI calculations.  In addition, they
can be handled with symbolic programs like Mathematica, which, while
computationally very slow, are an ideal environment for developing 
new methods and comparing the results with nearly analytical values.
Therefore, small model systems provide an ideal ``workbench'' for
testing new theories and algorithms. On the other hand, no method that
fails in the simplest case has hopes of succeeding in a realistic
calculation involving the more challenging Coulomb interaction with a
large number of electrons. Past experience has shown, in contrast,
that earlier SHDMC developments tested and developed in small
models~\cite{keystone} could be implemented easily in realistic cases
without additional complications~\cite{rollingstones}. 
Indeed, 
calculations using this method in QWALK~\cite{wagner09} 
reproduced the results obtained for the triplet state of He for low
magnetic fields\cite{jones99} starting from a random
linear combination of determinants constructed with the 
Hartree-Fock solutions {\it without} magnetic field. All 
electron calculations of atomic systems with tens of electrons are
currently under progress and will be published elsewhere\cite{elsewhere}.

DMC calculations with a realistic Coulomb interaction in periodic
systems require a supercell large enough to prevent unphysical image
interactions between periodic replicas of the electrons from
dominating the result. For the purpose of testing the method, however,
a model electron-electron interaction can be chosen, and the system can
be made as small as required for numerical convenience. For validating
the method, the Hamiltonian does not need to be strictly realistic;
however, one must solve the same Hamiltonian with SHDMC and an
established benchmark method (CI in this case).

The model studied in this section is related to the one considered in
Refs. \onlinecite{rosetta,keystone}, and \onlinecite{rockandroll} and consists of two
spinless electrons in a square of side $1$. However, instead of the
hard-wall boundary conditions used earlier, periodic boundary
conditions are set. 

{\it Basis expansion:} The ground state of the noninteracting system is
degenerate. Two states with zero total momentum can be constructed by
placing two electrons with opposite momenta ${\bf k}= \pm \pi {\bf
  \hat{\i}} $ or ${\bf k}= \pm \pi {\bf \hat{\j}} $.  The basis chosen to expand the
wave function is an antisymmetric combination of free-particle
solutions that satisfy periodic boundary conditions, which are plane
waves of the form
\begin{align}
\label{eq:basis}
e^{2\pi{\bf i}\left[  (n + 1/2) x + m y\right ]}
\end{align} 
where $|n+1/2|<6$ and $|m|<5$, which results in a two-body basis with
1516 functions.

The
confining potential and the interaction potential selected do not mix the
directions ${\bf \hat{\i}}$ and $ {\bf \hat{\j}}$. They are given by 
\begin{align}
\label{eq:perVR}
V({\bf R})= \; & 4 \pi^2 \left\{ 
\cos(2 \pi x_1)\! +\! \cos(2 \pi x_2)\! + \! 
              \cos(2 \pi y_1)\! + \! cos(2 \pi y_2) \right. \nonumber \\  
           &+ \left. \cos[2 \pi ( x_2 - x_1)] + \cos[2 \pi (y_2 - y_1)]) 
\right\}.
\end{align}
The first line of Eq.~(\ref{eq:perVR}) corresponds to an external
potential applied to electrons $1$ and $2$. The second line plays the
role of an interaction potential that depends on the difference
between the electronic coordinates.

 The Jastrow factor is set to zero to facilitate
the analytical calculation of the matrix elements of $V({\bf R})$, while
the kinetic energy is a diagonal matrix. 
The exact diagonalization of the Hamiltonian matrix is the CI
result.  For $\hbar=1$, the energy difference~\cite{units} between the noninteracting 
ground state and first excited states is $ 4 \pi^2$. Since
the interaction energy in Eq.~(\ref{eq:perVR}) is of the same order of
magnitude as the kinetic energy, the system is in the correlated
regime.

\subsection{Results and discussion}
\begin{figure}
\includegraphics[width=1.00\linewidth,clip=true]{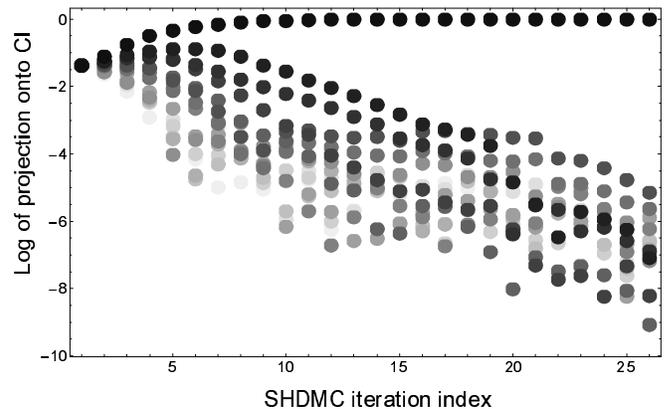}
\caption{ Logarithm of the projection  of the trial wave function into
  the lowest 16 eigenstates obtained with CI [$L_P(n) =
  ln|\langle\Psi_n^{CI}|\Psi_T(\ell\tau)\rangle|$] as a function of
  the SHDMC iteration index $\ell$. The results correspond to two
  electrons in the triplet state with periodic boundary conditions and
  a many-body Bloch phase $\phi({\bf R}) = \exp[{\bf i}({\bf k \cdot
    r}_1+{\bf k \cdot r}_2)]$ with ${\bf k}= 0.9 \pi({\bf
    \hat{\i}}+{\bf \hat{\j}}) $ (that is a twist angle of $1.8
  \pi$~\cite{tbc,fn:tbc}).  Darker symbols correspond to the projection
  with lower energy CI eigenstates. The initial trial wave function
  was a linear combination of the lowest 16 CI eigenstates with
  coefficients having the same modulus and a random complex phase.
  \label{fg:RPperiodic}}
\end{figure}

Figure \ref{fg:RPperiodic} shows the logarithm projection $L_P(n) =
ln|\langle\Psi_n^{CI}|\Psi_T(\ell\tau)\rangle|$ of the trial wave
function $|\Psi_T(\ell\tau)\rangle$ onto the $n$ eigenstate of the
full CI solution $|\Psi_n^{CI} \rangle$ as a function of the recursive
iteration index $\ell$.  The wave function is constrained by the basis
to have a many-body Bloch phase $\phi({\bf R}) = \exp[{\bf i}({\bf k \cdot
  r}_1+{\bf k \cdot r}_2)]$ with ${\bf k}= 0.9 \pi({\bf \hat{\i}}+{\bf
  \hat{\j}}) $ (that is a  twist angle of $1.8 \pi$~\cite{tbc,fn:tbc} both in 
the ${\bf \hat{i}}$ and ${\bf \hat{j}}$ directions). 

The initial trial wave function $|\Psi_T(0)\rangle$ was chosen
intentionally to be of poor quality to demonstrate the strength of the
method. The coefficients of $|\Psi_T(0)\rangle$ corresponded to a
linear combination of the first 16 full CI eigenstates:
$|\Psi_T(0)\rangle = \sum_n c_n |\Psi_n^{CI} \rangle$, where the
coefficients $c_n$ are complex numbers of modulus $1/4$ and a random
phase. Note that the initial trial wave function has no nodes but at
the coincidental points because is a linear combination with random phase
of different eigenstates of the non interacting Hamiltonian with 
different nodes.  The calculation was run for 200 walkers with
$\delta\tau=0.0004$ and $\tau=0.02$. The coefficients $\lambda_n$ were
sampled at the end of each sub-block of $k=50$ DMC steps. At first the
number of sub-blocks $M$ sampled before a wave function update was set
to $20$ and increased according to the recipe given in Ref.
\onlinecite{rockandroll} and briefly in Section \ref{sc:freephase}.
Therefore, the statistical error is reduced, and the number of basis
functions retained in the expansion increases over time. As a result
both the statistical error and the truncation error diminish, and the
wave function continues to improve.  The final iteration included $M =
600$ blocks.  The total optimization run cost $\approx 1.5\times 10^5$
DMC steps.

Figure \ref{fg:RPperiodic} shows in increasingly lighter shading
the results $L_P(n)$ corresponding to higher excited states.  All the
projections to the first 16 states start from the same value
[$-\ln(4)$] by construction. The algorithm, at first, increases the
projection of the lower energy states at the expense of the higher
ones (thus $L_P(n)$ approaches zero for low $n$), while the projections
with higher $n$ (in lighter gray) become smaller and their $L_P(n)$ is
increasingly negative. As the algorithm progresses further, the
projection on lower-energy excitations also starts to decay. Finally,
$L_P(n)$ becomes increasingly negative for all states 
except the ground state, which approaches zero. 

As the number of recursive iterations $\ell$ increases, the projection
onto highly excited states becomes negligible. The values obtained for
$L_p(n>0)$ are, therefore, dominated by statistical noise in the
sampling. On the right side of Fig. \ref{fg:RPperiodic}, the
convergence of the wave function is no longer limited by the initial
trial wave function but by the statistical noise. Statistical noise
introduces a projection into higher excited states by two mechanisms:
(i) the coefficients $\lambda_n$ of the trial wave function expansion
include random noise and (ii) the trial wave function develops
a projection into excited states because it is truncated depending on
the relative error of $\lambda_n$, which in turn depends on
$N_c$~\cite{keystone,rockandroll}.  Accuracy can  be increased only by
improving the statistics (increasing $M$ and $N_c$).  

The residual projection of the trial wave function
$|\Psi_T(\ell\tau)\rangle$ for iteration $\ell$ on the CI eigenstate
$|\Psi_n^{CI}\rangle$ is defined as
\begin{equation}
\label{eq:lrp}
L_{rp}^n=\ln\left(1-|\langle\Psi_n^{CI}|\Psi_T(\ell\tau)\rangle| \right).
\end{equation}
The final value for the residual projection for the calculation in
Fig. \ref{fg:RPperiodic} is below $-7$.  The value obtained for the
SHDMC energy is -31.842(13) as compared with a CI value of $-31.9486$.
However, the SHDMC wave function retains only $70$ coefficients in the
expansion, whereas the CI has $1516$. The FPDMC energy obtained with this
wave function was $-32.00(2)$.

The results shown in Figure \ref{fg:RPperiodic} demonstrate that the SHDMC
method with complex weights is able to correct both the phase and the
nodal structure of the trial wave function. SHDMC converges to the
ground-state even starting from a poor quality wave function with a
random phase.

\subsection{Many-body band structure}
Common electronic structure methods are based on a single-particle
picture, and the band structure is given by the evolution of the
energy as a function of the single-particle crystalline momentum. In
this case, in contrast, the energy of many-body states is a function
of the many-body Bloch phase $e^{ {\bf i} \left(\sum_j^{N_e} {\bf
      k\cdot r}_j\right) }$ or the twist angle~\cite{tbc,fn:tbc}.

\begin{figure}
\includegraphics[width=1.00\linewidth,clip=true]{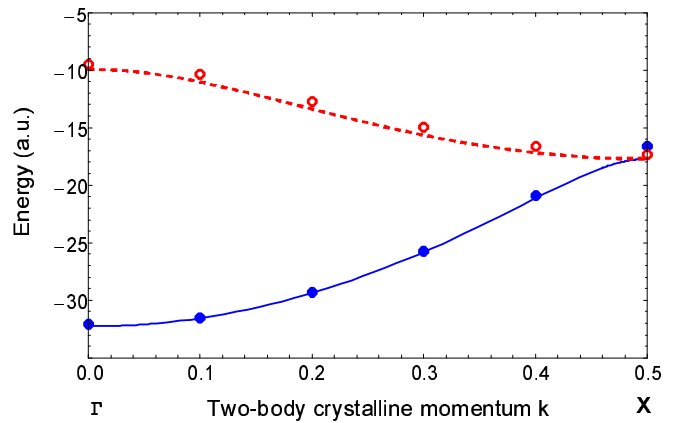}
\caption{ Many-body band structure of a periodic model system with 
  two electrons in the triplet state obtained with SHDMC (dots) and
  compared with CI (lines). The figure shows the energy of the ground
  state and the first excited state as a function of the global
  crystalline momentum ${\bf k}$ (i.e., the many-body Bloch phase or twist angle).
  \label{fg:bandstruct}}
\end{figure}

Figure \ref{fg:bandstruct} shows the many-body band structure for the
ground and first excited states as a function of the global crystalline
momentum ${\bf k}= k_x{\bf \hat{\i}}$ obtained for the same system
studied in Fig.~\ref{fg:RPperiodic}. The calculations were done using
the same parameters as in Fig.  \ref{fg:RPperiodic} described above.
The trial wave function for the ground state with ${\bf k}=0$ started
from a linear combination of the ground and first excited states of
the free-particle system with $\lambda_0=\lambda_1=1/\sqrt{2}$.  For
${\bf k} \ne 0$, the initial trial wave function for the ground state
was constructed using the Bloch part of the converged wave function
with smaller $|{\bf k}|$.  The initial trial wave function for the
first excited state for ${\bf k} = 0$ was constructed using a linear
combination including $\lambda_0$ and $\lambda_1$ orthogonal to the
ground state. The trial wave functions for the first excited states
for ${\bf k} \ne 0$ were constructed using the Bloch part of a
converged previous calculation with the closest value of ${\bf k}$ and using
the projector $\hat{P}$ to orthogonalize it with the ground state. CI
results are shown with lines for validation of the SHDMC results in
dots.  There is a very good agreement between the values obtained with
Quantum Monte Carlo and CI. In general, however, the Monte Carlo
values have a higher energy than the CI values. This is due to both the error in
the complex phase and the nodal error since the SHDMC wave function only retains  
$\approx70$ of the 1516 basis functions retained in the CI. The energy 
difference is reduced systematically as the algorithm progresses and more coefficients
$\lambda_n$ are retained in the trial wave function. 

\section{Ground and excited states with applied magnetic field}
\label{sc:magnetic}
\begin{figure}
\includegraphics[width=1.00\linewidth,clip=true]{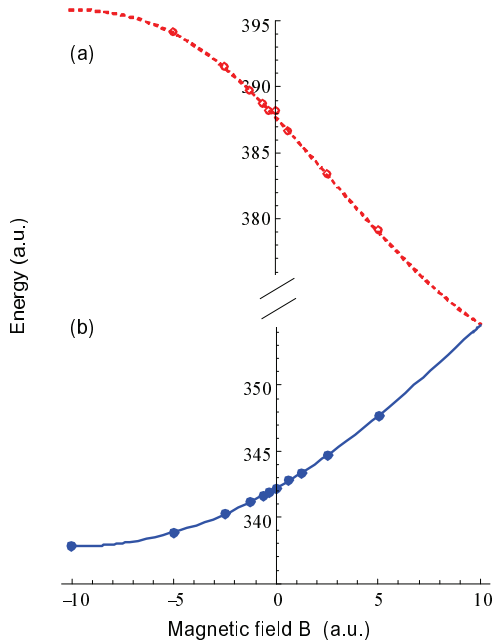}
\caption{\label{fg:magnetic} SHDMC results (dots) obtained for (a) the first excited state and (b)
  the ground state of a model system compared with CI results
  (lines) as a function of the magnetic field. The system consists of
  two electrons in a two-dimensional square in the triplet state. The
  results shown correspond to the ground and first excited states with
  $E$ symmetry that transform as $x+{\bf i}y$. The solution that
  transforms as $x-{\bf i}y$ can be obtained by changing the sign of
  $B$.}
\end{figure}
This section describes the results obtained with the generalization of
SHDMC (described in Section \ref{sc:freephase}) for the ground and the
first excited state of a model system with an applied magnetic field.
The results are compared with CI
calculations in the same model used in Refs.
\onlinecite{rosetta,keystone} and \onlinecite{rockandroll}.

\subsection{Model system with magnetic field}
Briefly, the lower energy eigenstates are found for two spinless electrons
moving in a two-dimensional square with a side length $1$ and a
repulsive interaction potential of the form $V({\bf r},{\bf
  r^{\prime}}) = 8 \pi^2 \gamma \cos{[\alpha
  \pi(x-x^{\prime})]}\cos{[\alpha \pi(y-y^{\prime})]}$ with $\alpha=
1/ \pi$ and $\gamma = 4$.  The many-body wave function is expanded in
functions $\Phi_n({\bf R})$ that are eigenstates of the noninteracting
system. The basis functions in $\{\Phi_n({\bf R})\}$ are linear
combinations of functions of the form $\prod_{\nu} \sin(m_{\nu} \pi
x_{\nu})$ with $m_{\nu} \le 7$.  Converged CI calculations were
performed to obtain a nearly exact expression of the lower energy
states of the system $\Psi_n({\bf R})= \sum_m a_m^n \Phi_m({\bf R})$.
The matrix elements involving the magnetic vector potential ${\bf A}$
(in the symmetric gauge) were calculated analytically using the
symbolic program Mathematica and were included in the CI Hamiltonian.  
The Jastrow factor was set to zero in the SHDMC run to
facilitate a direct comparison between CI and SHDMC results.

This paper reports results for the triplet case. In the absence of
a magnetic field, the triplet ground state is degenerate. Its orbital
symmetry corresponds to the E symmetry of the D$_{4}$ group. One of
the solutions with E symmetry transforms as $x$ and the other as
$y$. Under an applied magnetic field, the time reversal symmetry is broken,
and the $x$ and $y$ solutions are mixed.  Under a magnetic field, the
ground state can be expanded in a basis of functions that transform 
as $x \pm {\bf i} y$.  The energy of the $x- {\bf i} y$ solution can
be obtained from the energy of $x+ {\bf i} y$ by changing $B$ to $-B$.

\subsection{Results and discussion}
Figure \ref{fg:magnetic} shows energies of the ground state and first excited
state of the model system as a function of the magnitude
of the magnetic field $B$ (the curl of the vector potential ${\bf A}$). 
The calculations were run using $\delta \tau=0.00004$ and
$\tau=0.002$ and a total number of DMC steps of $10^5$ for each
calculated point. The calculation for the ground state started from
the noninteracting ground-state solution as a trial wave function.
The result obtained for $B = 0$ for the ground and first excited
states compared well with the ones obtained with the same Hamiltonian
in the triplet case reported~\cite{fn:errorCI} in Table I of
Ref. \onlinecite{rockandroll}. Note that in this case, the wave function is
complex and the coefficients have the freedom to be complex. Thus, in
contrast with Ref. \onlinecite{rockandroll}, where a real wave function was
enforced, here the phase was found within statistical error.

For $B \ne 0$, the time reversal symmetry is broken and so is the
degeneracy of the $x \pm {\bf i} y$ solutions. For higher (lower)
magnetic fields, the calculation began by using as the initial trial wave
function the one obtained previously with a lower (higher) magnetic
field.

The excited states were obtained using the method outlined in
subsection \ref{ssc:excited} and described in detail in Ref.
\onlinecite{rockandroll}. The lines show the CI results for reference.
The calculation for the first excited state with $B=0$ started from a
linear combination of the ground and first excited state of the
noninteracting system orthogonal to the interacting ground state
calculated earlier. The initial trial wave functions of the excited
states for $B \ne 0$ were taken from the previous calculations with
smaller $|B|$ (keeping the wave function orthogonal to the lower
energy states with the operator $\hat P$).  Clearly, Fig.
\ref{fg:magnetic}b shows good agreement between SHDMC and CI results
for the first excited state. 

Table \ref{tb:magnetic} summarizes the values obtained to construct
Fig.~\ref{fg:magnetic}. There is an excellent agreement in the
calculations obtained for the ground state using SHDMC and CI.  The
SHDMC energy values are, within error bars converged FPDMC results
indicating that the remaining convergence errors in the basis are
small. The agreement is less satisfactory for the excited states than
in the ground state (using the same computational time). It its clear
that the residual projections are much larger for the excited state
than for the ground.

An independent way to measure the quality of the wave function is the
logarithm of the variance of the modulus of the weights given by
\begin{equation}
\label{eq:logvar}
L_{var} = \ln{\sqrt{\frac{1}{N_c}\sum_{i,j} 
(|W_i^{k j}(k)| -1)^2 }} \; . 
\end{equation}
The variance of the weights does not deteriorate as much as the residual
projection for excited states, which might signal that the differences
in the wave functions originate because  CI and SHDMC minimize different
things using a truncated basis.~\cite{rockandroll}
\begin{table}
  \caption{\label{tb:magnetic} Comparison of the excitation energies obtained for the
    ground and the first excited state of a model system with two spinless 
    electrons and an applied magnetic field (see Fig. \ref{fg:magnetic}). $L_{rp}$ quantifies
    the overlap of the wave functions obtained with CI and SHDMC 
    [see Eq.~(\ref{eq:lrp})]. $L_{var}$ is the variance of the modulus of the walkers` 
weights [see Eq.~(\ref{eq:logvar})].}
\begin{ruledtabular}{|Ground State|}

\begin{tabular}{|c|l|c|c|c|c|}
\hline
 $B $ &  $E_0$ (SHDMC) & $E_0$ FPDMX  & $E_0$ (CI) & $L^0_{rp}$ & $L_{var}$ \\
\hline
-3.2 $\pi$ & 337.823   (13) & 337.820(7) & 337.821 & -9.9 & -4.7 \\
-1.6 $\pi$ & 338.877    (4) & 338.867(4) & 338.870 &-12.7 & -5.3 \\
-0.8 $\pi$ & 340.261    (7) & 340.256(5) & 340.256 &-12.9 & -5.7 \\
-0.4 $\pi$ & 341.143    (6) & 341.153(5) & 341.162 &-10.3 & -5.9 \\
-0.2 $\pi$ & 341.646   (11) & 341.662(6) & 341.667 &-13.6 & -6.0 \\
-0.1 $\pi$ & 341.931    (5) & 341.930(7) & 341.933 &-14.2 & -6.1 \\
 0.0 $\pi$ & 342.207    (7) & 342.206(5) & 342.208 &-11.9 & -6.1 \\
 0.2 $\pi$ & 342.771    (6) & 342.782(6) & 342.782 &-12.4 & -6.0 \\
 0.4 $\pi$ & 343.387    (5) & 343.392(4) & 343.390 &-10.8 & -6.0 \\
 0.8 $\pi$ & 344.696    (8) & 344.689(6) & 344.704 &-11.7 & -5.8 \\
 1.6 $\pi$ & 347.699    (8) & 347.684(5) & 347.697 & -9.4 & -5.2 \\
\hline
\end{tabular}
\end{ruledtabular}

\begin{ruledtabular}{|First Excited State|}
\begin{tabular}{|c|l|c|c|c|}
\hline
 $B $ &  $E_1$ (SHDMC) &$E_1$ (CI) & $L^1_{rp}$ & $L_{var}$ \\
\hline
 -1.6 $\pi$ &  394.161 (19) & 394.114 & -7.4 & -5.0 \\
 -0.8 $\pi$ &  391.532 (12) & 391.504 & -7.9 & -5.4 \\
 -0.4 $\pi$ &  389.744 (12) & 389.741 & -9.1 & -5.7 \\
 -0.2 $\pi$ &  388.786 (10) & 388.769 & -9.9 & -5.8 \\
 -0.1 $\pi$ &  388.253 (13) & 388.265 & -9.8 & -5.7 \\
  0.0 $\pi$ &  388.205 (44) & 387.750 & -5.7 & -5.0 \\
  0.2 $\pi$ &  386.697 (17) & 386.694 & -8.9 & -5.7 \\ 
  0.8 $\pi$ &  383.415 (14) & 383.407 & -8.9 & -5.4 \\
  1.6 $\pi$ &  379.159 (28) & 379.057 & -7.9 & -4.8 \\
\hline
\end{tabular}
\end{ruledtabular}
\end{table}

\section{Test with Coulomb interactions}
\label{sc:coulomb}
The calculations with Coulomb interactions were performed in the same 
system studied for the ground state in  Ref. \onlinecite{keystone} 
and for excited states in Ref. \onlinecite{rockandroll} but now with the additional
ingredient of an applied magnetic field. The more challenging 
triplet (antisymmetric) state was chosen for this study. 

The calculations were run with the same parameters and basis as in
Fig.~\ref{fg:magnetic} and Table~\ref{tb:magnetic} but with a Coulomb
interaction potential of the form $V({\bf r},{\bf r^{\prime}}) = 20
\pi^2 /|{\bf r-r\prime}|$. Since the average of the Coulomb
interaction is much larger than the single-particle energy differences,
the system is in the highly correlated regime.
\begin{table}
\caption{\label{tb:coulomb} SHDMC and FPDMC energies as a function of an applied magnetic 
  field for a model system with
  two electrons in a triplet state in a square box with Coulomb interactions. The quality of the
  wave function is measured by
  $L_{var}$ 
  [see Eq.~(\ref{eq:logvar})].
}
\begin{ruledtabular}
\begin{tabular}{|c|l|c|c|c|}
State &  $B$   & SHDMC & PFDMC&$L_{var}$ \\
\hline
0 & -1.60 $\pi$ & 401.65 (2) & 401.67(4)&-4.1 \\
0 & -1.26 $\pi$ & 401.80 (3) & &-4.2 \\
0 & -0.80 $\pi$ & 401.92 (3) & 401.87(4) &-4.2 \\
0 & -0.40 $\pi$ & 403.50 (6) & 402.39(7) &-3.4 \\
0 & -0.20 $\pi$ & 402.97 (4) & 402.60(5) &-4.1 \\
0 & 0.00        & 402.76 (4) & 402.58(3)  &-4.0 \\
0 & 0.40 $\pi$  & 403.23 (2) & 403.20(3)  &-4.6 \\
0 & 0.80 $\pi$  & 403.87 (3) & 403.73(3)  &-4.2 \\
0 & 1.26 $\pi$  & 404.93 (6) & &-3.7 \\
0 & 1.60 $\pi$  & 405.54 (9) & 405.16(4)  &-3.8 \\
\hline 
1 & -0.40 $\pi$ & 465.37 (10)& &-3.0 \\
1 & -0.20 $\pi$ & 468.55 (7) & &-3.5 \\
1 & 0.00        & 454.39 (8) & &-3.4 \\
1 & 0.40 $\pi$  & 451.76 (8) & &-3.2 \\
2 & -0.40 $\pi$ & 486.89 (7) & &-3.2 \\
\end{tabular} 
\end{ruledtabular}
\end{table}

Table \ref{tb:coulomb} displays the values obtained for the model
system with Coulomb interactions for the ground state and some
excitations as a function of the magnetic field. The quality of the
wave function is characterized by the logarithm of the variance of the
modulus of weights given by Eq.~(\ref{eq:logvar}).  Note that the
variance of the weights increased when Coulomb interactions are
considered when compared with the case of the model interaction. This is due to 
the Coulomb singularity and the lack of a Jastrow factor.  While the
variance of the weights is larger in the Coulomb case, the quality of
the wave function improves from one SHDMC recursive iteration to
the next (see below).

\subsection{Improvement of the wave function's node and phase with SHDMC }
\begin{figure}
\includegraphics[width=1.00\linewidth,clip=true]{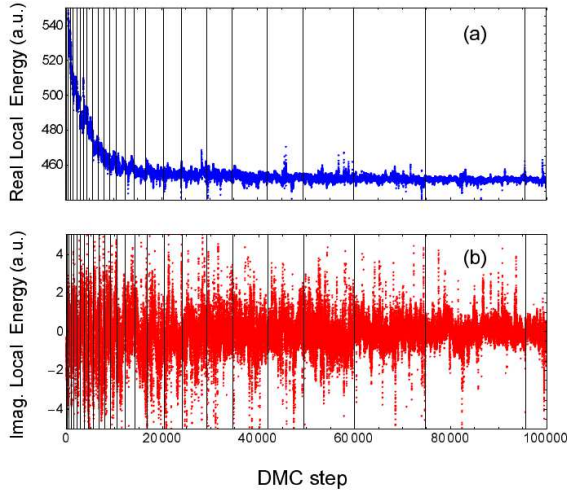}
\caption{\label{fg:coulomb} (Color online) (a) Average of the real part
  of $E_L({\bf R})$ obtained with 200 walkers as a function of the DMC
  step.  The results correspond to the first excited state that
  transforms as $x+{\bf i} y$ with $E$ symmetry of the group $D_{4}$
  of two electrons with Coulomb interactions in a square box with an
  applied magnetic field~\cite{units} $B = 0.4 \pi $. (b) Average of
  the imaginary part of $E_L({\bf R})$ as a function of the DMC step
  for the same case.  The vertical lines mark the end of the SHDMC
  block when the wave function is updated. }
\end{figure}

Figure \ref{fg:coulomb} shows the evolution of the real (a) and the
imaginary (b) parts of the local energy $E_L({\bf R})$ as a function
of the DMC step for the first excited state of two electrons in a
square box with an applied magnetic field of $0.4 \pi$.  The
calculations started with a trial wave function with two nonzero
coefficients chosen to be orthogonal to the ground state calculated
earlier.  It can be clearly seen in Fig. \ref{fg:coulomb}(a) that as
the SHDMC algorithm progresses, the real part of the local energy is
quickly reduced and stabilized at the first excited state energy.  The
imaginary part of the $E_L({\bf R})$ should be zero for an eigenstate;
otherwise, the divergence of the current is 
nonzero~\cite{ortiz93}. In SHDMC this strong condition is satisfied only
as the number of recursive iterations, the number of configuration
sampled $N_c$, and the size of the basis $N_b$ retained in the wave
function tend to infinity. Figure \ref{fg:coulomb}(b), however,
clearly shows that the variance of the raw data obtained for
$Im[E_L({\bf R})]$ is reduced as the SHDMC algorithm progresses. This
is a clear indication of improvement of the phase of the wave
function.

\begin{figure}
\includegraphics[width=1.00\linewidth]{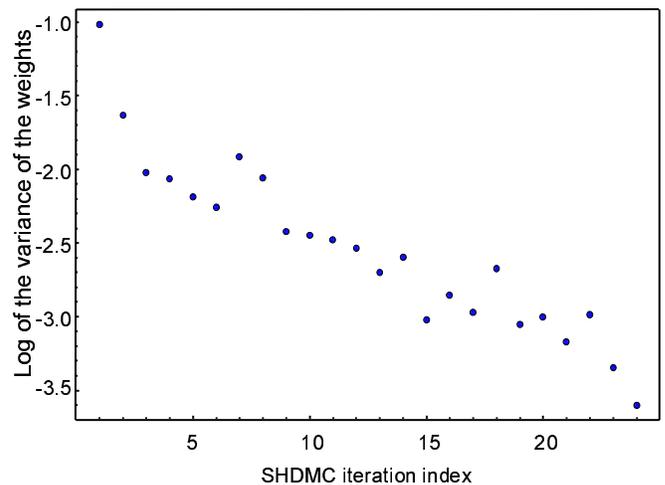}
\caption{(Color online) Logarithm of the variance of the 
  modulus of weights as a function of the SHDMC block index $\ell$. The results correspond to the
  first excited state with  $B = 0.4 \pi $ and Coulomb interactions 
(the run shown in Fig.~\ref{fg:coulomb}). 
\label{fg:variance}}
\end{figure}

Figure \ref{fg:variance} shows the evolution of 
logarithm of the variance of the weights [see Eq.~(\ref{eq:logvar})] 
as a function of the SHDMC block index (the number of wave function updates).
The reduction in weight variance is a clear indication of 
convergence of the trial wave function towards an eigenstate of the 
Hamiltonian~\cite{rockandroll}.

\section{Summary and perspectives}
\label{sc:conclusion}
A method that allows the calculation of the complex amplitude of a
many-body wave function has been presented and validated with model
calculations. An algorithm that finds the complex wave function is
essential for any study of many-body Hamiltonians with periodic
boundary conditions or under external magnetic fields.  The method
converges to nearly analytical results obtained for model systems
under applied magnetic fields or periodic boundary conditions, with an
accuracy limited only by statistics and the flexibility of the wave
function sampled. 

It is found that for some eigenstates, the ones 
where the phase is a scalar function of ${\bf R}$, there is a {\it special} gauge
transformation in which wave function is real. For this class of eigenstates the
original proof of convergence of SHDMC applies. For complex wave functions 
some fermionic eigenstates may not have nodes. In the latter case, as in the case
of bosons\cite{rockandroll,reatto82}, the convergence
of SHDMC is not affected since the wave function can evolve everywhere. 

This new approach goes beyond both fixed-phase DMC~\cite{ortiz93} and
SHDMC~\cite{keystone,rockandroll,rollingstones}. As in the real wave
function version of SHDMC, the method is recursive and the
propagation to infinite imaginary time is achieved as the number of
iterations increases. As in FPDMC, the walkers evolve under an
effective potential that incorporates the gradient of the phase of
the trial wave function and the vector potential of the magnetic
field.  But in this new algorithm, in contrast to FPDMC, the complex
amplitude of the wave function is free to adjust both its modulus and
its phase.  After each iteration, the trial wave function is improved
following a short time evolution of an ensemble of walkers.  These
walkers follow the equation of motion of a generalized importance
sampling approach.  Unlike previous attempts, the walkers carry a
complex weight resulting from eliminating the
fixed-phase constraint in the time evolution of the mixed probability
density.  The modulus of the weight can be used to calculate real
observables, such as the energy. The phase of the weight of the walkers
is used to improve the phase of the trial wave function in the
following iteration. As in earlier versions of SHDMC, the modulus of
the weights is also used to improve, simultaneously, the node if there is 
any and the phase of the 
trial wave function.

This free amplitude SHDMC method can be used to calculate not
only the ground state but also low energy excitations~\cite{rockandroll}
within a DMC context. Comparisons with nearly
analytical results in model systems demonstrate that the new approach
converges to the many-body wave function of systems with applied
magnetic fields or with periodic boundary conditions for low energy
excitations.

This recursive method finds a solution to ``the phase problem'' and,
if there is any, finds the node at the same time. The many-body
wave function can be used, in principle, to calculate any
observable.  However, in very large systems, when convergence with the
size of the wave function basis cannot be fully achieved, a standard
fixed-phase calculation should be performed as a final step to obtain a
more accurate energy.

{\it Scaling and cost:} An analysis of the minimum cost required to
determine the node and the phase has to take into account the number
of independent degrees of freedom of the Hilbert space.  Arguably, no
method could scale better than linear in the number of {\it
  independent} degrees of freedom of the problem studied; otherwise,
some degrees of freedom would be {\it dependent} from each other.  A
real-space expansion of the many-body wave function with fixed
resolution $L_R$ is ideal for counting independent degrees of freedom.
The resolution $L_R$ can be connected to the energy cutoff of the
excitations in a multideterminant expansion. For a complex wave
function, each point in the many-body space ${\bf R}$ has two
independent degrees of freedom (modulus and phase).  If the volume of
a system is proportional to the number of electrons $N_e$, its size
scales as $L \approx \alpha N_e^{1/3}$ (where $\alpha $ is of the
order of the Bohr radius $a_B$).  Taking into account the $N_e !$
permutations of identical particles, one finds $(L/L_R)^{3 N_e}/N_e! $
independent degrees of freedom for each spin channel to determine the
phase $\phi({\bf R})$.  Thus, the number of independent degrees of
freedom scales as $\exp\{N_e(3 \log(\alpha/L_R))+1)\}$. The node
$S_T({\bf R})$, if there is any, requires one less dimension (which, if the
nodal surface is not too convoluted, could reduce the number of
degrees of freedom by only up to a factor $(L/L_R)$). Since the number
of independent degrees of freedom of the phase increases exponentially
with $N_e$, for a fixed resolution $L_R$ one cannot find the phase
with an algorithm polynomial in $N_e$.

This generalization of the SHDMC method, though tested in small
systems, is targeted to be used in large systems.  The numerical cost
of SHDMC scales linearly with the number of independent degrees of
freedom of the phase per recursive step. However, the number of
independent degrees of freedom (i.e, the size of the basis expansion)
should increase exponentially with the number of electrons $N_e$ for a
fixed resolution. The accuracy of SHDMC is limited by the size of the
basis sampled, the statistical error, and the number of recursive
iterations. The number of recursive steps required increases
if the product between $\tau$ and the lowest energy excitations is
small. The SHDMC method can be used in combination with other
optimization approaches to accelerate convergence in that limit.

The scaling of the cost of exact diagonalization methods such as CI is at
least quadratic with the number of degrees of freedom. Often a CI
calculation is used to preselect a multideterminant expansion to be
improved within a VMC context before a final FPDMC run.  An advantage
of SHDMC is that it incorporates the Jastrow in the sampling of the
coefficients. Thus SHDMC might be more efficient than a CI filtering
for large systems.  The linear scaling of SHDMC suggests that it could
be the method of choice to optimize the wave function phase and nodes for
calculations in periodic solids.

The optimization of many-body wave functions with current in periodic
boundary conditions is now possible. Therefore, the new method can be
used as a tool to perform transport calculations including many-body
effects. The calculation of systems with an applied magnetic field is
challenging, even in the case of small molecules and atoms and
particularly so when the magnetic field, the many-body interactions,
and the kinetic energy are of the same order of
magnitude~\cite{jones96}. The calculations reported in this paper,
though in a simple model, suggest that the method can be applied to
the study of molecular or atomic systems in that difficult regime.

Our recent successful application of the ground-state algorithm for
real wave functions~\cite{keystone} to molecular
systems~\cite{rollingstones} supports the idea that this
generalization of SHDMC can also be useful for real ab-initio
calculations beyond model systems. The implementation of the algorithm
in state-of-the-art DMC codes has been done. Initial results in atomic
systems show that the many-body wave function improves, which is shown
by a reduction of the average local energy, the energy variance and
the variance of the imaginary contribution to the total energy.

\acknowledgments The
author would like to thank J. McMinis for an introduction to the
fixed-phase approximation and P. R. C. Kent, and G. Ortiz for a
critical reading of the manuscript. 
The author also thanks M. Bajdich
for sharing all electron calculations in atomic systems
using this method as supplemental material for the referees prior
publication.  Research sponsored by the Materials Sciences \&
Engineering Division of the Office of Basic Energy Sciences U.S.
Department of Energy.


\begin{thebibliography}{99}
\bibitem{fn:basis} A complete basis in the symmetric space can be used
  to calculate bosons. Other symmetries of the wave function can be
  enforced with the selection of the basis~\cite{rockandroll}.
\bibitem{fn:taup} Since $\tau^\prime = \tau \ell$ is essentially an iteration index,
it is omitted in the trial wave function phase and amplitude for clarity.
\bibitem{fn:discont} The position of the cut in the complex
plane of the Riemann surface into sheets is arbitrary. Therefore, the
discontinuities in the gradient and the effective potential are no physical if
they can be removed changing $n$.
\bibitem{fn:tbc} The term ``twist angle'' was introduced~\cite{tbc} to
  avoid confusion with other usages of the term ``phase''. Here the
  term ``many-body Bloch phase'' is also mentioned since the Bloch
  theory is well understood outside the many-body field.
\bibitem{units} The energy unit is $\hbar^2/(2 m L^2 )$ and the magnetic 
unit is given by $e/(c \hbar
m^{(1/2)}) $.
\bibitem{fn:errorCI} There was a small error in the CI calculations reported in
Ref. \onlinecite{rockandroll}.
\end{thebibliography}
\end{document}